\documentclass[a4paper,twocolumn,showpacs,amsmath, amssymb,nofootinbib,aps,floatfix]{revtex4}
\usepackage{epsfig}
\input psfig.sty 

\usepackage{bm}									
\usepackage{dcolumn}								
\usepackage{graphicx}								
\usepackage{hyperref}

\newcommand{\no}{\mbox{$n_{e_o}$}}

\begin{document}

\title{Cosmography with the Sunyaev-Zeldovich effect and X-ray data}

\author{R. F. L. Holanda$^{1,2}$\footnote{E-mail: holanda@uepb.edu.br}}

\author{J. S. Alcaniz$^3$\footnote{E-mail: alcaniz@on.br}}

\author{J. C. Carvalho$^3$\footnote{E-mail: carvalho@dfte.ufrn.br}}

\address{$^1$Departamento de F\'{\i}sica, Universidade Estadual da Para\'{\i}ba, 58429-500, Campina Grande - PB, Brasil}

\address{$^2$Departamento de F\'{\i}sica, Universidade Federal de Campina Grande, 58429-900, Campina Grande - PB, Brasil}

\address{$^3$Departamento de Astronomia, Observat\'orio Nacional, 20921-400, Rio de Janeiro - RJ, Brasil}

\date{\today}

\begin{abstract}
Cosmography provides a direct method to map the expansion history of the Universe in a model-independent way. Recently, different kinds of observations have been used in cosmographic analyses, such as SNe Ia and gamma ray bursts measurements, weak and strong lensing, cosmic microwave background anisotropies, etc. In this work we examine the prospects for constraining cosmographic parameters from current and future measurements of galaxy clusters distances based on their Sunyaev-Zeldovich effect (SZE) and X-ray observations. By assuming the current observational error distribution, we perform Monte Carlo simulations based on a well-behaved parameterization for the deceleration parameter to generate samples with different characteristics and study the improvement on the determination of the cosmographic parameters from upcoming data. The influence of galaxy clusters (GC) morphologies on the $H_0- q_0$ plane is also investigated.

\end{abstract}


\maketitle

\section{Introduction}

Cosmic acceleration is one of the most important issues of modern theoretical physics (see, e.g., Sahni \& Starobinski 2000; Peebles \& Ratra 2003; Padmanabhan 2003; Alcaniz (2006) and Li et al. (2011) for recent reviews). However, more than a decade after its discovery, the physical mechanism behind this phenomenon remains unknown.  In the context of Einstein's general relativity (GR), this result implies either the existence of a new field, the so-called dark energy, or that the matter content of the universe is subject to dissipative processes (Lima \& Alcaniz, 1999; Chimento et al. 2003).

In order to investigate this phenomenon in a model-independent way cosmographic approaches have been successfully applied (see, e.g., Turner \& Riess 2002; Bamba et al. 2012). This approach is independent on the gravity theory and the matter-energy contents filling the Universe and provides  an interesting way to map and explore the expansion history of the universe. Besides SNe Ia observations, other cosmological observables have been used to employ a kinematic description of the Universe as, for instance, cluster strong lenses (D'Aloisio \& Natarajan 2011), SNe Ia plus cosmic microwave background (CMB) and Baryon Acoustic Oscillation (BAO) (Santos et al. 2011; Demianski et al. 2012), BAO plus observational Hubble data and gamma ray bursts (Xu \& Wang 2011), weak gravitational lensing plus gamma ray bursts (Wang \& Dai 2011). 

On the other hand, the promising technique of measuring galaxy clusters distances based on their Sunyaev-Zeldovich effect (SZE) and X-ray observations have not been fully explored in the kinematic context. Our goal in this paper is to derive cosmographic bounds on the $H_0- q_0$ plane by using angular diameter distance (ADD) from galaxy clusters. To this end we use a parametric approximation of the deceleration parameter along the cosmic evolution, given by $q(z) = q_0 + q_1z/(1 + z)$. In order to ensure a period of structure formation during the matter-dominated era, we also make use of the asymptotic value of $q(z)$ at high redshift, which reduces the above parameterization to an one-parameter function. Initially, we use  25 ADD of galaxy clusters as compilled by De Filippis et al. (2005). Since these authors obtained the ADD by using two morphological description for galaxy clusters (elliptical and spherical $\beta$ models) we also explore the influence of the morphology on the results. By assuming the 
observational error distribution of the De Filippis et al. sample, we also perform Monte Carlo simulations to generate samples with different sizes and study the expected improvement on the determination of the cosmographic plane $H_0- q_0$ from upcoming data. In what follows, we outline the main assumptions for our analysis and discuss the main results.

\section{Galaxy clusters distances} 

The so-called SZE is a  small distortion of the Cosmic
Microwave Background (CMB) spectrum provoked by the inverse
Compton scattering of the CMB photons passing through a population
of hot electrons (Sunyaev \& Zel'dovich 1972). The effect is
proportional to the electron pressure integrated along the line of
sight, i.e., to the first power of the plasma density. Another important physical phenomenon occurring in the intracluster medium is the X-ray emission that occurs primarily through thermal bremsstrahlung. The X-ray
surface brightness $S_X$ is proportional to the integral along the
line of sight of the square of the electron density. Briefly, in the context of these two phenomena one may
consider the different electronic density dependencies and
evaluate the angular diameter distance of the galaxy cluster such that (Silk \& White 1978; Cavaliere \& Fusco-Fermiano 1979),
\begin{equation}
{\cal{D}}(z)\propto \frac{(\Delta
T_{0})^{2}\Lambda_{eH0}}{(1+z)^4
S_{X0}{T_{e0}}^{2}}\frac{1}{\theta_{c}},
\end{equation}
where $S_{X0}$ is the central X-ray surface brightness, $T_{e0}$ is
the central temperature of the intracluster medium, $\Lambda_{eH0}$
is the central X-ray cooling function of the intracluster medium.
$\Delta T_0$ is the central decrement temperature, and
$\theta_{c}$ refers to a characteristic scale of the cluster along
the line of sight whose exact meaning depends on the assumptions
adopted to describe the galaxy cluster morphology (Carlstrom, Holder \& Reese 2002). This technique of measuring distances is completely independent of other techniques and it can be used to measure distances at high redshifts directly. Recently this technique has been applied to a fairly large number of clusters (Reese et al. 2002; Jones et al. 2005; De Filippis et al. 2005; Bonamente et al. 2006) -- see also Mason et al. 2001; Cunha, Marassi \& Lima 2007; Lima, Holanda \& Cunha 2010; Holanda, Cunha \& Lima 2012, Holanda, Cunha, Marassi \& Lima 2012 for current cosmological constraints using ADD from galaxy clusters observations.

It worth mentioning that, in order to estimate the angular distance of a galaxy cluster from its SZE/X-ray observations, one needs to add some complementary assumptions about its geometry. In the last decade, many studies about the intracluster gas and dark
matter distribution in galaxy clusters have been limited to the standard spherical geometry (Reiprich \& Boringer 2002; Bonamente et al. 2006; Shang, Haiman \& Verdi 2009). The importance of the intrinsic geometry of the cluster has been emphasized by many authors (Fox \& Pen 2002; Jing \& Suto 2002; Plionis, Basilakos \&  Ragone-Figueroa 2006; Sereno et al. 2006) and the standard spherical geometry has been severely questioned, since Chandra and XMM-Newton observations have shown that clusters usually exhibit an elliptical surface brightness. { It is worth mentioning that the first determination of the intrinsic three-dimensional (3D) shapes of galaxy clusters was presented by Morandi et al. (2010) by combining X-ray, weak-lensing and strong-lensing
observations. Their methodology was applied to the galaxy clusterMACS J1423.8+2404 and they found a triaxial galaxy
cluster geometry with DM halo axial ratios $1.53 \pm 0.15$ and $1.44\pm 0.07$ on the plane of the sky and along the line of sight,
respectively.} In recent papers, the elliptical morphology  was also shown to be in agreement with the validity of the so-called cosmic distance duality relation $D_L=(1+z)^2D_A$, where $D_L$ is the luminosity distance (Holanda, Lima \& Ribeiro 2010, 2011, 2012; Nair, Jhingan, \&  Jain 2011 ).

\section{Cosmography}

Let us now assume that the Universe is spatially flat as suggested by current WMAP measurements (Hinshaw et al. 2012). In this case, the ADD in the FRW metric is defined by ($c=1$),
\begin{eqnarray}\label{eq:dLq}
D_A &=& (1+z)^{-1}H^{-1}_{0}\int_0^z {du\over H(u)} = \frac{(1+z)^{-1}}{H_0} \nonumber \\
&& \,\, \int_0^z\, \exp{\left[-\int_0^u\, [1+q(u)]d\ln
(1+u)\right]}\, du,
\end{eqnarray}
where  $H(z)=\dot a/a$ is the Hubble parameter and the deceleration parameter $q(z)$ is defined by

\begin{eqnarray}\label{qz}
q(z)\equiv -\frac{a\ddot a}{\dot a^2} = \frac{d H^{-1}(z)}{ dt} -1.
\end{eqnarray}

In order to proceed further, we adopt in our analysis the following parametrization for the deceleration parameter: 
\begin{eqnarray}\label{pqz}
q(z) = q_0 + q_1z/(1 + z), 
\end{eqnarray}
where $q_0$ is the current value of the deceleration parameter and in the infinite past $q(\infty) = q_0 + q_1$. This parametric form was inspired by one of the most popular parameterizations of the dark energy equation of state (Chavallier \& Polarski 2001; Linder 2003) and, although very simple, seems to be flexible enough to mimic the $q(z)$ behavior for a wide class of accelerating models. However, for most of the viable cosmological scenarios, we expect the Universe to be matter-dominated at early times (i.e., after the radiation dominance), which implies $q =1/2$. Thus, in order to ensure this past dark matter-dominated epoch whose existence is fundamental for the structure formation process to take place, we assume the constraint $q(z >> 1) = 1/2$ (Santos, Carvalho \& Alcaniz 2011). In this case, $q_0 + q_1= 1/2$ and Eq. (3) becomes 
\begin{equation}
q(z) = (q_0 + \frac{z}{2})\frac{1}{1+z},
\end{equation}
which has the advantage of reducing our analysis to a two-parameter fitting, $q_0$ and $H_0$. In what follows, we perform our analysis using this latter expression.

\begin{figure}
\label{Fig}
\psfig{figure=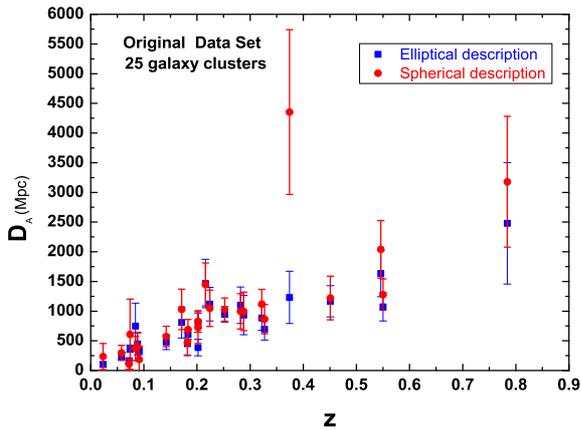,width=3.2truein,height=2.3truein}
\hskip 0.1in
\caption{25 angular diameter distances from galaxy clusters. The red filled circle and the blue filled square correspond to elliptical and spherical description for the same galaxy clusters sample.}
\end{figure}

\begin{figure*}
\label{Fig}
\centerline{
\psfig{figure=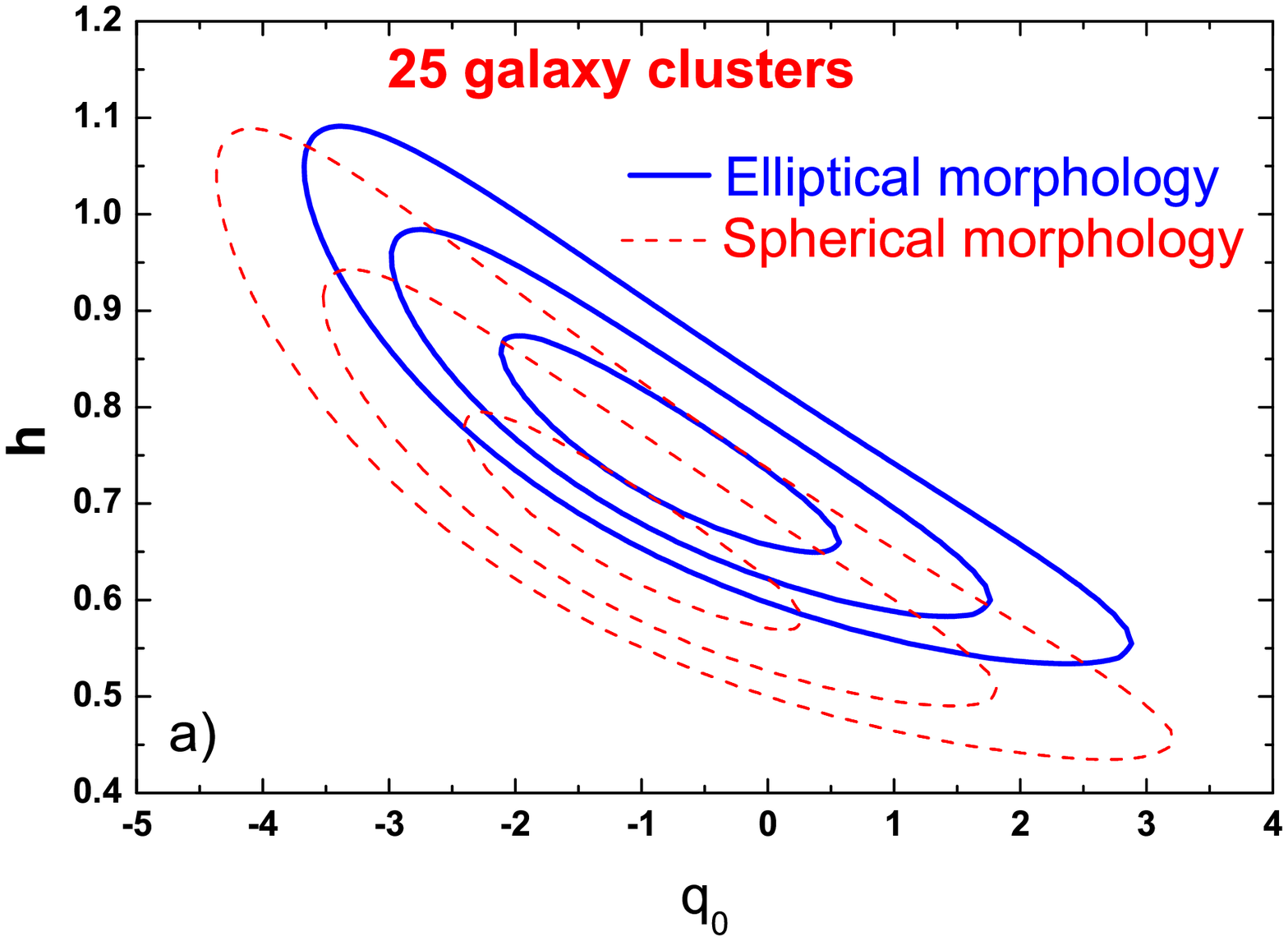,width=2.3truein,height=2.3truein}
\psfig{figure=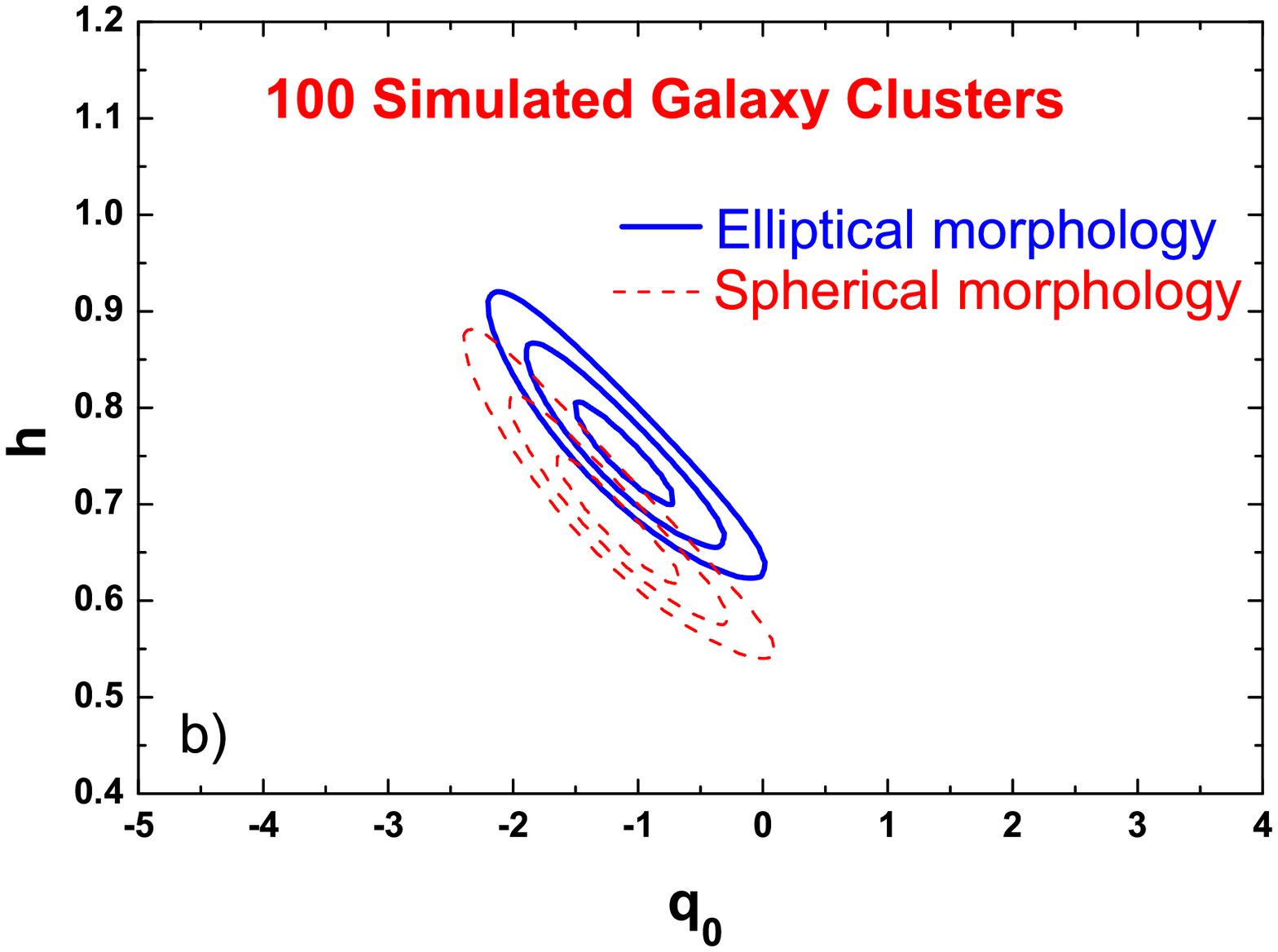,width=2.3truein,height=2.3truein}
\psfig{figure=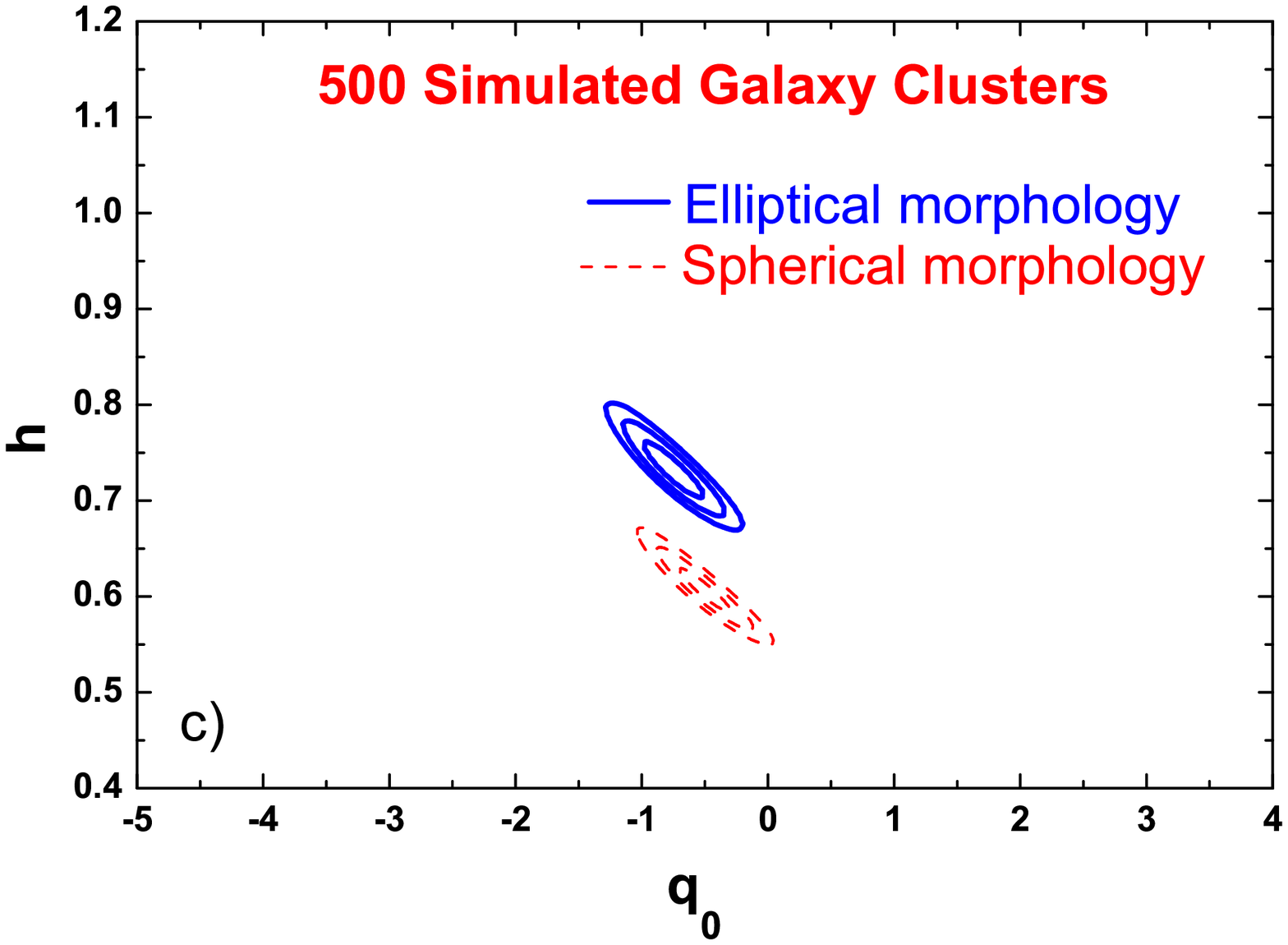,width=2.3truein,height=2.3truein}
\hskip 0.1in}
\caption{Confidence contours on the plane $h - q_0$ for three data samples of ADD measurements of galaxy clusters and different morphologies. {\it{Left)}} 25 observational data of De Filippis et al. (2005). {\it{Middle)}} 100 simulated data points and {\it{Right)}} 500 simulated data points.}
\end{figure*}

\section{Galaxy clusters observations}

\subsection{Current samples}

 In order to derive cosmographic bounds on the epoch of cosmic acceleration we use the  angular diameter distances of galaxy clusters from the De Filippis et al. (2005) compilation. This sample is composed of 25 data in redshift range $0.023 \leq z \leq 0.8$. These authors re-analyzed archival X-ray data with XMM-Newton and Chandra satellites of two samples for which combined X-ray and SZE
analysis have already been reported using an isothermal spherical
$\beta$-model. One of the samples, compiled previously by Reese et al.\ (2002),
is a selection of 18 galaxy clusters distributed over the redshift
interval $0.14 < z < 0.8$. The another one (Mason et al.\ 2001) has 7 clusters from the X-ray limited flux sample of
Ebeling et al.\ (1996). 

{  Besides the standard isothermal spherical $\beta$ model, de Filippis et al. (2005)  used 
isothermal elliptical $\beta$-models to obtain $D_A(z)$ measurements for these cluster  samples. As discussed in this latter reference, the choice of circular rather than elliptical $\beta$ model does not affect the resulting central surface brightness or Sunyaev-Zeldovich decrement and the slope $\beta$ differs slightly between the two models. However, significantly different values for the core radius are obtained. De Filippis et al. (2005) found that $\theta_{ell} =
\frac{2e_{proj}}{1+e_{proj}}\theta_{circ}$, where $e_{proj}$ is the
axial ratio of the major to the minor axes of the projected
isophotes, $\theta_{circ}$ and  $\theta_{ell}$ are the angular core radius of the spherical and elliptical $\beta$ models, respectively. As is well known $D_A \propto \theta^{-1}_c$, so that  angular diameter
distances obtained by using an isothermal spherical $\beta$-model
are overestimated compared with those from the elliptical $\beta$-model. By using these two approaches we can directly compare the results of different  morphological assumptions.} The two samples of De Filippis et al. (2005) are shown in Fig. 1. The blue squares and filled red circles with the associated error bars  stand for the same galaxy clusters but  described by elliptical and spherical models, respectively. 

\subsection{Simulations}

We have run a series of Monte Carlo simulations to generate synthetic samples of $D_A(z)$ for both the spherical and elliptical model. As a fiducial model, we used the best-fit values of $q_0$ and $H_0$ obtained from statistical analysis with the current observational data, i.e., $q_0=-0.85^{+1.35}_{-1.25}$ and $H_0 = 75 \pm 10$ $\rm{km.s^{-1}.Mpc^{-1}}$ (elliptical morphology) and $q_0=-1.17^{+1.40}_{-1.30}$ and $H_0 = 67.0 \pm 13$ $\rm{km.s^{-1}.Mpc^{-1}}$ (spherical morphology). Using these  values, we generated a number of numerically simulated ADD measurements. In order to estimate the error bars of the distance values, as well as the dispersion of the errors, we made a detailed study of the current data. Because of the relatively small number of observed clusters, a bootstrap analysis was carried out to make the numerical simulations more reliable. We assumed that at a given redshift, the values of $D_A$ are normally distributed around the fiducial value, the error bars being drawn from a normal 
distribution as well. Samples containing 100 and 500 clusters were then generated. The statistical study of these synthetic data was then carried out and is presented in next section.

\begin{figure*}
\label{Fig}
\centerline{
\psfig{figure=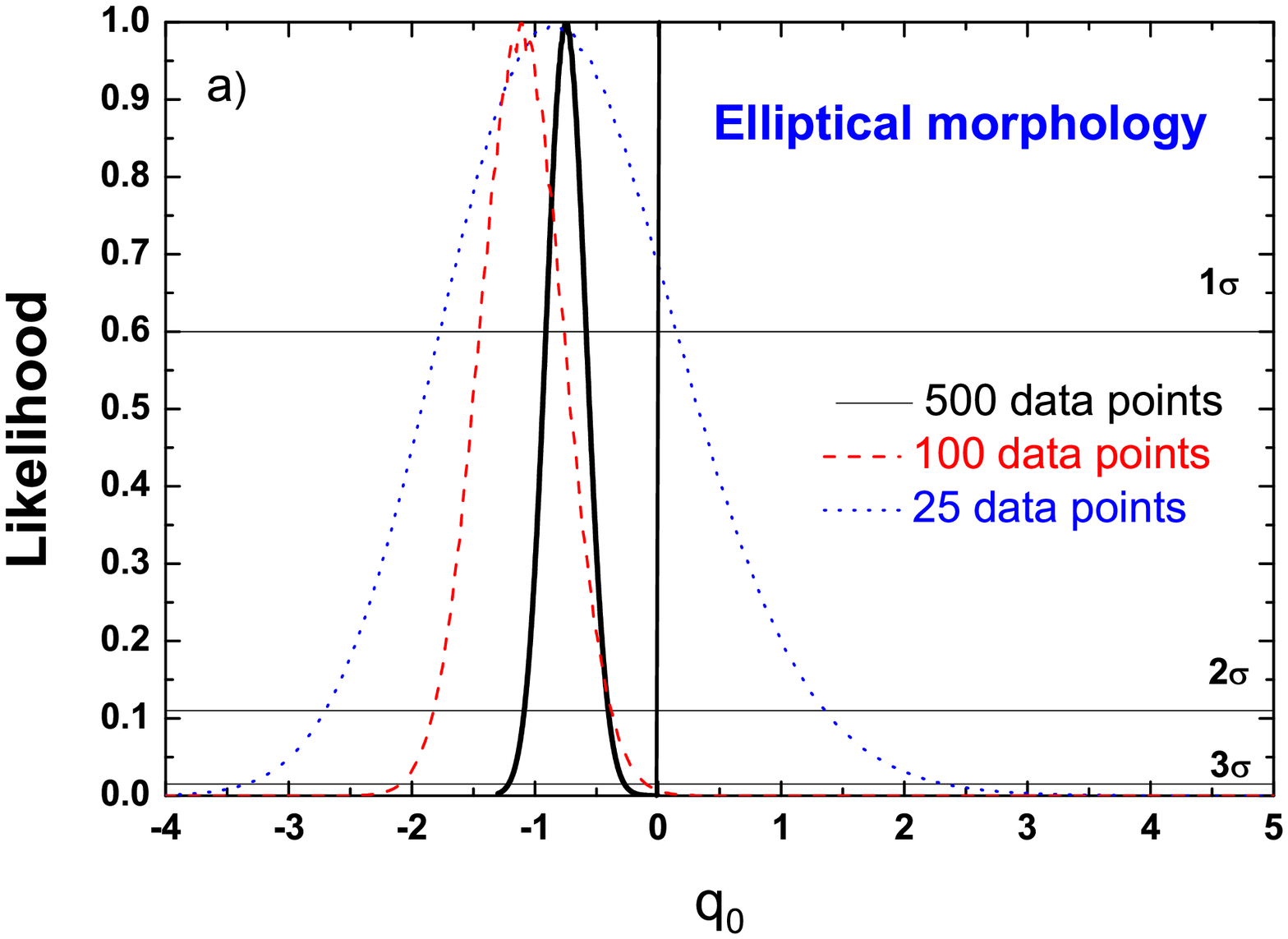,width=3.0truein,height=2.3truein}
\hskip 1.5cm
\psfig{figure=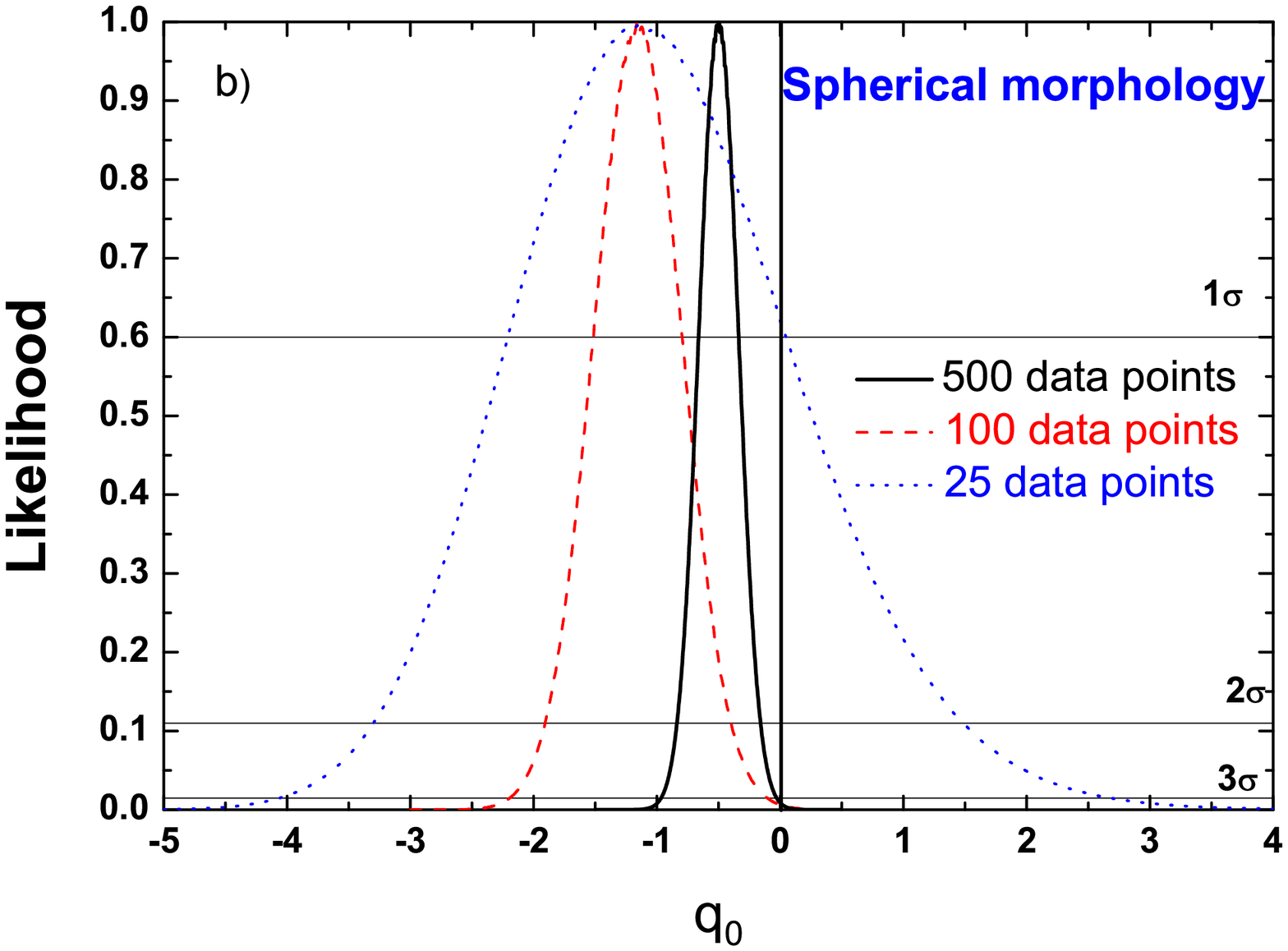,width=3.0truein,height=2.3truein}
\hskip 0.1in}
\caption{{\it{Left:}} Normalized likelihood for $q_0$ from the three samples discussed in the text marginalizing over the Hubble parameter and considering a elliptical cluster morphology. {\it{Right:}} The same as in the previous panel for spherical morphology. }
\end{figure*}

\begin{figure*}
\label{Fig}
\centerline{
\psfig{figure=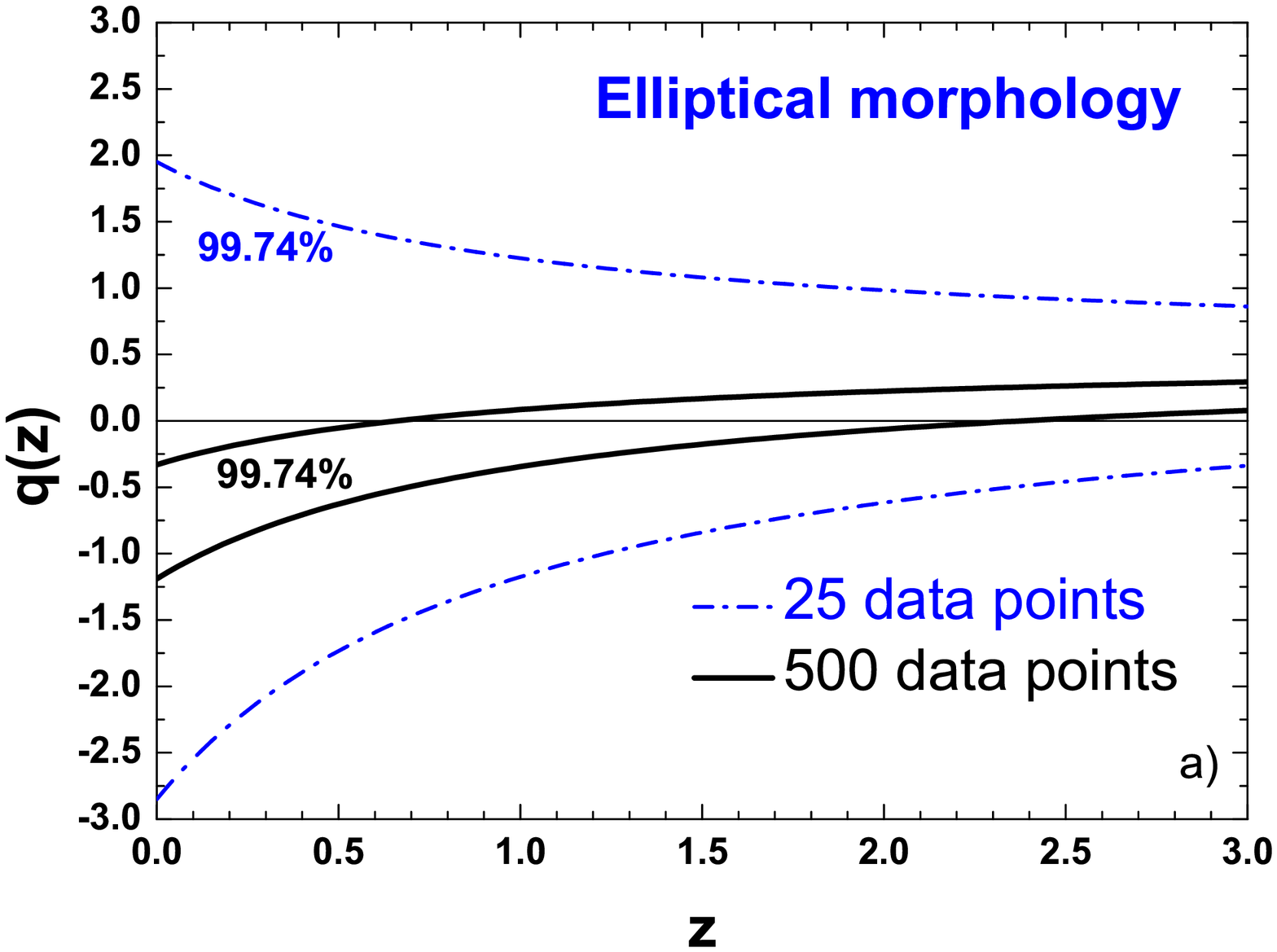,width=3.0truein,height=2.3truein}
\hskip 1.5cm
\psfig{figure=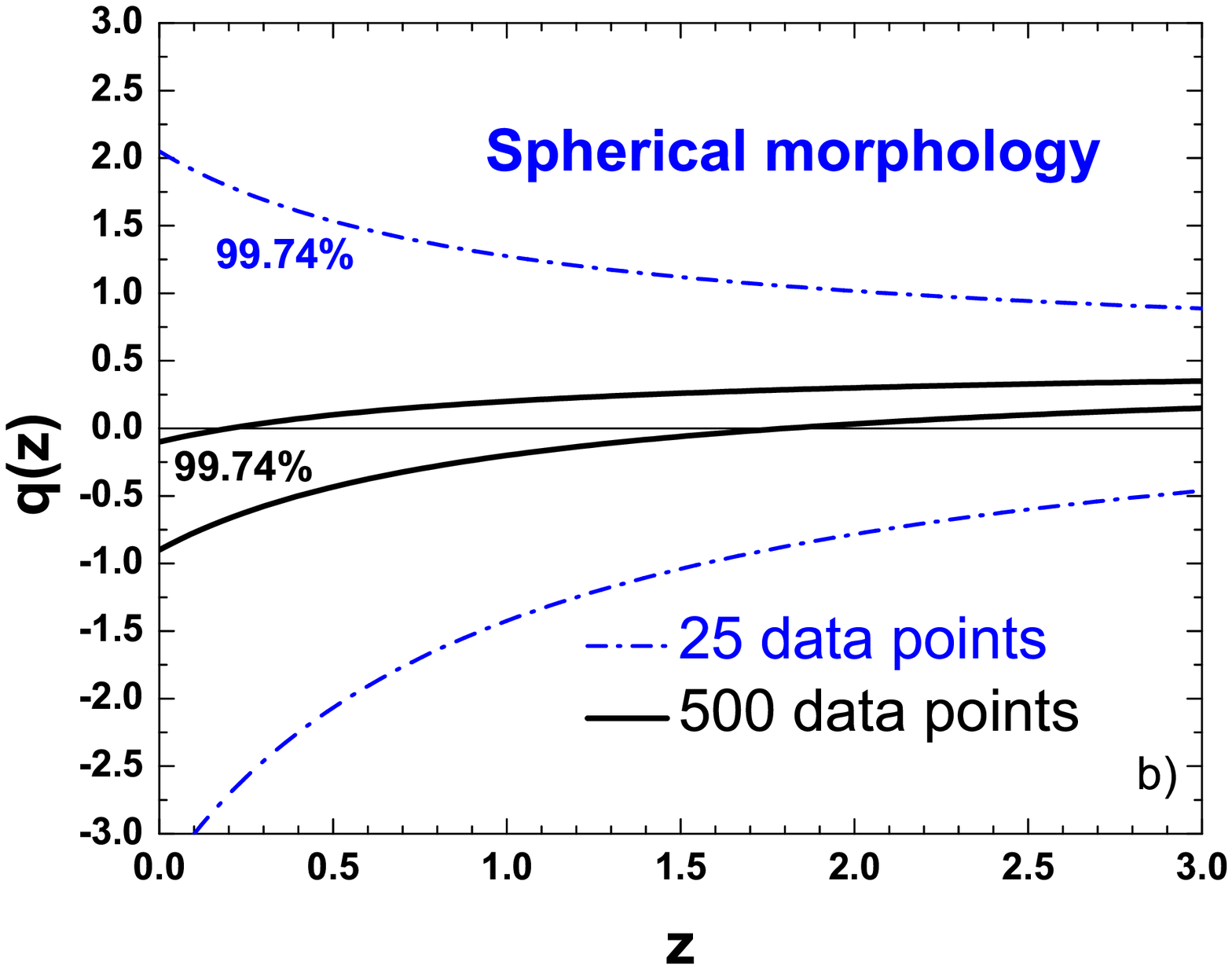,width=3.0truein,height=2.3truein}
\hskip 0.1in}
\caption{{\it{Left:}} 2$\sigma$ bounds on the evolution of the deceleration parameter from observed (dot-dashed) and simulated (solid) samples assuming a elliptical cluster morphology. {\it{Right:}} The same as in the previous panel for spherical morphology.}
\end{figure*}

\section{Analysis and Results}

In what follows, we perform a $\chi^2$ fit over the $(q_0,H_0)$
plane. In our analysis we use a maximum likelihood that can be
determined by a $\chi^2$ statistics,
\begin{equation}
\chi^2(z|\mathbf{p}) = \sum_i { (D_A(z_i; \mathbf{p})-
{\cal{D}}_{Ao,i})^2 \over \sigma_{{\cal{D}}_{Ao,i}}^2},
\end{equation}
where ${\cal{D}}_{Ao,i}$ is the observational ADD,
$\sigma_{{\cal{D}}_{Ao,i}}$ is the statistical uncertainty in the individual
distance, $D_A$ is ADD as given by Eq. (2) and the complete set of parameters is given by
$\mathbf{p} \equiv (q_0,H_0)$. { Note that 
\begin{equation}
{\cal{D}}_{Ao,i} \propto \frac{\Delta T_{CMB}^2 \Lambda_{ee}}{S_X T_e^2}.
\label{eq-DA}
\end{equation}
Thus, ${\cal{D}}_{Ao,i}$ is proportional to $\Delta T_{CMB}^2$ and $T_e^{3/2}$ (since
$\Lambda_{ee} \propto T_e^{1/2}$) and distance determinations are strongly dependent on the accuracy of the
SZE decrement and X-ray temperature measurements (a detailed discussion of statistical and systematic errors can be found in Bonamente et al. 2006). Here, however, we emphasize that the most common statistical error contributions  to the ADD of galaxy clusters are: SZE point sources $\pm 8$\%, X-ray background $\pm 2$\%, Galactic N$_{H}$ $\leq \pm 1\%$, $\pm 15$\% for cluster asphericity, $\pm 8$\% kinetic SZ and for CMBR anisotropy $\leq \pm 2\%$ (see  table 3 in Bonamente et al. 2006).  The total statistical error in ${\cal{D}}_{Ao,i}$ are calculated by combining the individual statistical uncertain  in quadrature. As a matter of fact, one may show that typical statistical errors amount for nearly $20$\%, in agreement with other analyses (Mason et al.\ 2001; Reese et al.\ 2002, 2004). On the other hand, the estimates for systematic effects are SZE calibration $\pm 8$\%, X-ray flux calibration $\pm 5$\%, radio halos $+3$\%, and X-ray temperature calibration $\pm 7.5$\%. Many of the systematics can be approached and reduced through improved observations (Morandi et al. 2013, Hasler et al. 2012).}

   \begin{table}[ph]
\caption{$1\sigma$ estimates on the cosmographic parameters} 
{\begin{tabular} {@{}ccc@{}} Data points  & $\sigma_{q_0}/q_0$ & $\sigma_{H0}/H_0$  \\ \hline \hline
&elliptical morphology& \\ \hline
25 (observed) &$1.58$& $0.14$ \\ 
100 (simulated) &$0.38$&$0.08$ \\ 
500 (simulated) &$0.31$&$0.04$ \\ \hline
&spherical morphology & \\ \hline
25 (observed)  &$1.15$&$0.19$ \\
100 (simulated) &$0.60$&$0.13 $ \\
500 (simulated) &$ 0.42$&$0.05 $
\\ \hline
\end{tabular} \label{ta1}}
\end{table}

In Figs. 2a - 2c we show the constraints on the plane ($q_0,H_0$) by using exclusively the SZE and X-ray observations. In Panel 2a, we show the 68.3\%, 95.4\% and 99.7\% contours for the De Filippis et al. (2005) samples with the solid and dashed lines corresponding to elliptical and spherical models, respectively. For these data and irrespective of the model adopted, we find very loose bounds on the cosmographic parameters $q_0$ and $H_0$ with both accelerating ($q_0<0$) and decelerating ($q_0>0$)  universes being compatible at high confidence level.

Our results show that, even considering the current observational error distribution, an increase in the number of data points decreases considerably the uncertainty on the parametric plane $q_0 -H_0$ (see Table I).  It is also worth mentioning that the constraints on $q_0$ are insensitive to the galaxy clusters morphology, with the analyses with 100 and 500 data points ruling out a decelerating universe at $3\sigma$ level.

To quantify the improvement  on the determination of the cosmographic parameters, we define our figure-of-merit as FoM $\propto 1/\sigma_{q_0} \sigma_{H_0}$ and find:
$$
\rm{FoM_{500}} = 3.91 \times \rm{FoM_{100}} = 23.13 \times \rm{FoM_{25}}  \quad (\rm{Elliptical})\; , 
$$
$$
\rm{FoM_{500}} = 7.56 \times \rm{FoM_{100}} = 25.92 \times \rm{FoM_{25}}  \quad (\rm{Spherical})\; .
$$

In Figs. 3a and 3b we plot the $q_0$ likelihoods  for the two morphological descriptions when the Hubble parameter is marginalized over. Clearly, a larger sample (100 and 500 data points) with the same characteristics of the currently observed one can rule out a current decelerated universe. For the sake of completeness, we also plot in Figs. 4a and 4b a $2\sigma$ reconstruction of the $q(z)$ function using both the current observational data (25 points) and 500 simulated data points. Differently from the current observations, the 500 simulated data points shows a switch from a decelerated to an accelerating phase.

\section{Conclusions}

In this work we have discussed cosmographic bounds from measurements of the angular diameter distance of galaxy clusters based on their Sunyaev-Zeldovich effect and X-ray observations. By using a parametric approximation of the deceleration parameter along the cosmic evolution given by Equations (2) and (5), we have also explored the influence of galaxy clusters morphology on the estimates of $H_0$ and $q_0$. 

Two different analyses have been performed. First, we have derived bounds on the plane $H_0 - q_0$ from the current sample of 25 ADD measurements for which the Sunyaev-Zel'dovich effect and X-ray surface brightness maps were fitted by an isothermal spherical $\beta$ model and an isothermal elliptical $\beta$ model. Irrespective of the morphology adopted, we have found very loose bounds on the space $H_0 - q_0$, with both accelerating ($q_0<0$) and decelerating ($q_0>0$)  universes being compatible at high confidence level. In addition, we have performed Monte Carlo simulations of the ADD of galaxy clusters based on the current observational error distribution (De Filippis et al. 2005) to study the dependence of the cosmographic bounds with the size of the sample.  We have shown that, even keeping the current statistical observational uncertainty, an increase in the number of data points increases considerably the figure-of merit for the cosmographic plane $H_0 - q_0$. These results, therefore, highlight the cosmological interest in ADD measurements of galaxy clusters from the Sunyaev-Zeldovich effect and X-ray observations and show that upcoming measurements may become complementary or even competitive with other cosmological probes. We emphasize that SZE and X-ray de
termined distances are independent of the extragalactic distance latter and do not rely on clusters being standard candles or rulers.

The authors thank CNPq, INCT-A and FAPERJ for the grants under which this work was carried out.

\label{lastpage}

\begin{thebibliography}{99}

\bibitem{alcaniz} Alcaniz, J. S., 2006, Braz.\ J.\ Phys.\  {\bf 36}, 1109 [astro-ph/0608631].
\bibitem{bamba}Bamba, K., Capozziello, S.,  Nojiri, S., Odintsov, S. D. 2012, to appear in Astrophysics and Space Science
\bibitem{Boname06} Bonamente, M., et al. 2006, ApJ, 647, 25
\bibitem{Carlstrom02} Carlstrom, J. E., Holder, G. P., \& Reese, E. D. 2002, ARA\&A, 40, 643
\bibitem{caval} Cavaliere, A., \& Fusco-Fermiano, R. 1978, A\&A, 667, 70
\bibitem{chavalier} Chavallier, M. \&  Polarski, D. 2001,  Int. J. Mod. Phys. D 10, 213
\bibitem{chi}Chimento, L. P., Jakubi, A. S., Pavón, D. \& Zimdahl  W. 2003, PRD, 67, 083513 
\bibitem{CML07}  Cunha, J.\ V., Marassi, L., \& Lima, J.\ A.\ S. 2007, MNRAS, 379, L1 [astro-ph/0611934]
\bibitem{Natarajan} D'Aloisio, A. \& Natarajan, P. 2011, MNRAS, 411, 1628
\bibitem{DeFilippis05} De Filippis, E., Sereno, M., Bautz, M.\ W., \&
        Longo G.\ 2005, ApJ, 625, 108
\bibitem{Marek} Demianski, M., Piedipalumbo, E.,  Rubano, C. \& Scudellaro, P. 2012, accepted in MNRAS       
\bibitem{ebelin} Ebeling, H., et al.\ 1996, MNRAS, 281, 799
\bibitem{FoxPen02} Fox, D. C. \&  Pen, U. L. 2002, ApJ, 574, 38
\bibitem{hasler}{ Hasler N. et al. 2012, ApJ, 748, 12} 
\bibitem{wmap}Hinshaw, G. et al. 2012,  submitted to ApJS (WMAP Team)
\bibitem{holapjl}{ Holanda, R.\ F.\ L., Lima, J.\ A.\ S.,\ \& Ribeiro,
        M.\ B.\ 2010, ApJL, 722, L233 [arXiv:1005.4458]}
\bibitem{Holandab}Holanda, R.~F.~L., Lima, J.~A.~S., \& Ribeiro, M.~B.\ 
        2011a, A\&A Letters, 528, L14 [arXiv:1003.5906]
\bibitem{holanda2011}{ Holanda, R.\ F.\ L., Lima, J.\ A.\ S.\ \& Ribeiro, M.\ B.\ 2012, A\&A, 538, A131}   
\bibitem{holapjl} Holanda, R. F. L., Marassi, L., Cunha, J. V., \& Lima, J. A. S. 2012, JCAP, 02, 035
\bibitem{Holanda} Holanda, R. F. L., Cunha, J. V., \& Lima, J. A. S. 2012, GRG, 44, 501
\bibitem{jing}Jing, Y. P. \& Suto, Yasushi 2002, ApJ, 574, 538
\bibitem{Jones05} Jones, M. E., et al. 2002, MNRAS, 357, 518
\bibitem{LHC}Lima, J. A. S., Holanda, R. F. L., \& Cunha, J. V. 2010, AIP Conf. Proc., 1241, 224 
\bibitem{LI}Li, M., Li, X-D \& Wang, S. 2011, Commun. Theor. Phys., 56,  525
\bibitem{Linder}Linder, E. V. 2003, Phys. Rev Lett. 90, 091301  
\bibitem{lima}Lima, J. A. S. \& Alcaniz, J. S. 1999, A\&A, 348, 1        
\bibitem{Mas01} Mason, B.\ S., et al. 2001, ApJ, 555, L11
\bibitem{morandi}{ Morandi, A., Nagai, D. \& Cui, W. 2013, 431, 1240}
\bibitem{djain} Nair, R., Jhingan, S., \&  Jain, D.\ 2011, JCAP, 05, 023
\bibitem{Pad}Padmanabhan, T. 2003, Phys. Rep., 380,  235 
\bibitem{peebles}Peebles, P. J. \& Ratra, B. 2003, Rev. Mod. Phys., 75,  559
\bibitem{Plionis06} Plionis, M.,  Basilakos, S., \&  Ragone-Figueroa, C. 2006, ApJ, 650, 770
\bibitem{rb02}Reiprich, T. H., \& Bohringer, H. 2002, ApJL, 567, 716
\bibitem{Reese02}  Reese, E.\ D., et al. 2002, ApJ, 581, 53
\bibitem{Reese04}  Reese, E.\ D.\ 2004, in Measuring and Modeling the
        Universe, ed. W.\ L.\ Freedman (CUP) p. 138 [astro-ph/0306073]
\bibitem{santos} Santos, J., et al. 2008 Phys. Lett. B 669, 14
\bibitem{santosb} Santos, B.; Carvalho, J. C. \& Alcaniz, J. S. 2011, Astroparticle Physics, 35, 17
\bibitem{sereno}	Sereno, M.; De Filippis, E.; Longo, G. \& Bautz, M. W. 2006, ApJ, 645, 170
\bibitem{shang}	Shang, C.,  Haiman, Z. \& Verde, L. 2009, MNRAS, 400, 1085
\bibitem{sherwin}Sherwin, B. D., et al. 2011, PRL, 107, 021302
\bibitem{SilWhi78} Silk, J., \& White, S. D. M. 1978, ApJ, 226, L103
\bibitem{Sarni}Sahni, V. \& Starobinsky, A. 2000, IJMPD, 9, 373
\bibitem{SunZel72} Sunyaev, R.\ A., \& Zel'dovich Ya.B.\ 1972,
        Comments Astrophys.\ Space Phys., 4, 173
\bibitem{turner}  Turner,M. S., \&  Riess, A. G. 2002, ApJ, 569, 18        
\bibitem{wang}Wang, F. Y. \& Dai, Z. G. A\&A, 2011, 536, A96        
\bibitem{Lixin} Xu, L., \&  Wang, Y. 2011, PLB, 702 114       
\end{thebibliography}
\end{document}